\documentstyle[prl,aps,preprint]{revtex}

\begin{document}
\tighten

\title{INFRARED RENORMALONS AND POWER CORRECTIONS
IN DEEP-INELASTIC SUM RULES\thanks
{This work is supported in part by funds provided by the U.S.
Department of Energy (D.O.E.) under cooperative agreement
\#DF-FC02-94ER40818.}}

\author{Xiangdong Ji}

\address{Center for Theoretical Physics \\
Laboratory for Nuclear Science \\
and Department of Physics \\
Massachusetts Institute of Technology \\
Cambridge, Massachusetts 02139 \\
{~}}

\date{MIT-CTP-2381 ~~~HEP-PH /9411312 ~~~
           Submitted to: {\it Nucl. Phys. B} ~~~ November 1994}

\maketitle

\begin{abstract}%
Infrared renormalons and $1/Q^2$ power corrections in
deep-inelastic sum rules are studied. The renormalization
of operators with power divergence are discussed.
The higher-twist terms in the operator product
expansion are shown to account for the residual
soft contributions survived from the
Kinoshita-Lee-Nauenberg type of cancellation
in Feynman diagrams. The presence of some degree of
arbitrariness in the twist separation
allows one to define the most convenient
higher-twist operators suitable for a particular
non-perturbative method. The discussion
is focused on the Bjorken sum rule, for which the
$1/Q^2$ corrections are considered on a lattice.
\end{abstract}
\pacs{xxxxxx}

\narrowtext

\section{Introduction}

Deep-inelastic sum rules---Bjorken sum rule \cite{BJ}, Gross-Llewellyn Smith
sum rule \cite{GLS}, Ellis-Jaffe sum rule \cite{EJ}, and a whole
zoo of similar sum rules \cite{YND}---are fascinating things
to study, both from theoretical
and experimental points of view. These sum rules form one of
the clearest tests of perturbative and non-perturbative
Quantum Chromodynamics (QCD), and
tests of our ability to solve the theory as well. Some of the sum rules
were derived before QCD was advanted, in current algebras
that the asymptotic QCD turns out to respect up to logarithms.
In QCD, they can be derived
using the technique of Wilson's operator product expansion (OPE)
or its equivalent. The imprints of QCD in these sum rules are
radiative corrections, which can be calculated perturbatively,
and the higher-twist corrections that involve high-dimensional hadron
matrix elements. This paper is intended to address the issue on the
separation of these two effects and some related aspects in the
OPE.

In order to be specific, I will center my discussion on the
Bjorken sum rule, which presently is under active experimental
study \cite{SLAC}. The sum rule is for the nucleon's spin-dependent
structure function $g_1(x, Q^2)$ and the QCD version reads \cite{SV},
\begin{equation}
  \int^1_0(g_1^p(x, Q^2)-g_1^n(x, Q^2))dx
= {1\over 6}C_2(\alpha_s(Q^2))g_A - {4\over 27}C_4(\alpha_s(Q^2))
   {\langle O^\mu_4  \rangle \over Q^2}+ ...,
\end{equation}
where $p$ and $n$ label proton and neutron, respectively, and
$g_A=1.257$ is the neutron $\beta$-decay constant.
$C_i(\alpha_s(Q^2)$ are the coefficient functions
that can be calculated in perturbation theory as power series
in $\alpha_s(Q^2)$,
\begin{equation}
   C(\alpha_s) = \sum_{n=0}^{\infty}c_n \alpha^n_s .
\label{pert}
\end{equation}
The second term in the right-hand-side of Eq. (1) is the twist-four
correction in which $\langle O_4^\mu\rangle$ is the hadron matrix
element,
\begin{equation}
     \langle PS|O^\mu_4|PS \rangle
      = 2S^\mu \langle O^\mu_4 \rangle ,
\end{equation}
with
\begin{equation}
     O^\mu_4 = \bar u g\tilde F^{\mu\alpha}\gamma_\alpha u
         - \bar d g\tilde F^{\mu\alpha}\gamma_\alpha d ,
\end{equation}
where $|PS\rangle$ is the nucleon state with momentum $P^\mu$
and polarization $S^{\mu}$
($\langle PS|PS\rangle = 2P^0(2\pi)^3\delta^3(0)), $
and $u$, $d$, and $F^{\mu\nu}$ are
quark and gluon fields. Both $C_4$ and $\langle O^{\mu}_4
\rangle $ depend on a factorization (renormalization)
scale $\mu^2$, however, the dependence cancels out in the
product. For simplicity, I have neglected the target
mass corrections and the twist-three contribution \cite{JU}, both
of which are present at order $1/Q^2$, but inessential
for the following discussion. The ellipsis denotes still
higher-twist contributions that are suppressed by
more powers of $1/Q^2$.

It is commonly accepted that the
first (twist-two)
and the second (twist-four) terms as well as
higher-order terms in $1/Q^2$ in Eq. (1) are separately
gauge invariant and finite, and hence the twist-separation
is well-defined. Indeed, the leading-order
coefficient function $C_2(\alpha_s(Q^2)$ is often
calculated in terms of the virtual Compton amplitude
for a zero-momentum quark state in dimensional
regularization \cite{LV}, in which perturbative
higher-twist matrix elements vanish identically, and
the non-perturbative higher-twist matrix elements
are known to be calculable, for instance, in lattice QCD
by subtracting the corresponding perturbative matrix
elements \cite{LAT}, or in the QCD sum rule method \cite{BAL}.
It turns out, however,
this traditional view on the twist-separation
is at best formal, and one in practice is forced
to provide more details on the individual
contribution. Eventually, the twist expansion
has the similar degree of arbitrariness as the
factorization of the coefficient
functions from the operator matrix elements in
individual term of the OPE \cite{CSS}. This surprising
feature of the twist-expansion in deep-inelastic sum rules
was first recognized and studied, to the author's knowledge,
by A. Mueller \cite{MUL1}, although similar studies in vacuum current
correlation functions had begun much early and the results are
much more well-known \cite{HOO,PAR,DAV,MUL2}.

This paper grows out of attempts to understand the $1/Q^2$
effects in the deep-inelastic sum rules \cite{JU2} and Mueller's
paper on ambiguity of the higher-twist contribution \cite{MUL1}.
Since I believe most of the physicists
working in deep-inelastic scattering are unaware of this issue,
I intend to be pedagogical in my presentation. Therefore,
the discussions in the paper are often mixed
with results that are already known
in the case of vacuum correlation functions.
I hope the references are sufficiently clear so the
reader has no confusion about what is new and what is not.

In section II, I discuss the infrared (IR) renormalons in the coefficient
functions of the Bjorken sum rule. I start with
an empirical observation and a calculation of a bubble-chain
diagram to convince the reader that the perturbation
series in fact diverges. Then I state some general results known
about the properties of
the renormalon singularities. In section III, I consider
the renormalization of operators with power-like divergence
and explain why the multiplicative
renormalization of log divergence
cannot define the operators uniquely.
The conclusion is that
{\it pure} non-perturbative operators in general do not exist
and higher-twist operators
depend on handling of the perturbative contributions.
In section IV, I derive a cut-off version of the OPE by
separating the soft and hard contributions to
Feynman diagrams. The result shows that the coefficient
functions can be defined without IR renormalons,
but depending on a momentum cut-off, and
the higher-twist operators depend on the same cut-off
due to perturbative power contributions.
However, the sum of the twist expansion is cut-off-independent.
In section V, I argue that there exists a large degree of
freedom in choosing the higher-twist operators. According to
this, I define a twist-four operator for the Bjorken sum rule
suitable for lattice calculations. The corresponding twist-two
coefficient function is subsequently evaluated. The final section
contains the discussion on the practical significance of
the $1/Q^2$ corrections in the Bjorken sum rule,
and summary and conclusion of this paper.

\section{Infrared renormalons in coefficient functions }

The IR renormalons, in the Bjorken sum rule, begin
with the calculation
of the coefficient function $C_2(\alpha(Q^2))$ in dimensional
regularization and minimal subtraction scheme (${\overline{\rm MS}}$).
According to text books \cite{MUTA}, the coefficient functions $C_i$ are
the coefficients of local operators in Wilson's OPE,
\begin{equation}
   \int d^4\xi e^{iq\cdot \xi}
     T\left[J_\mu^{\rm em}(\xi)J_\nu^{\rm em}(0)\right] = -2i
   \epsilon^{\mu\nu\alpha\beta}{ q_\alpha \over Q^2}\left[
    {1\over 3}C_2(\alpha_s(Q^2)) J_{5\beta}(0)
     - {8\over 27Q^2} C_4(\alpha_s) O_{4\beta} + ... \right],
\label{ope}
\end{equation}
where $J_{5\beta} = \bar u\gamma_\beta\gamma_5 u-\bar d\gamma_\beta
\gamma_5 d$ is the isovector axial current whose nucleon matrix
element is $g_A$. Other quantities are defined following
Eq. (1). Since Eq. (\ref{ope}) is an operator identity,
it is valid when sandwiched in any state. The canonical way of
calculating $C_2(\alpha_s(Q^2))$ is to apply Eq. (\ref{ope})
to a single quark state with zero momentum. It is easy to see that
the higher-twist matrix elements vanish in
dimensional regularization, and $C_2(\alpha_s)$ is
just the ''ratio" of the quark Compton amplitude to the quark
matrix element of the axial current. In fact, this is exactly
what has been down in Ref. \cite{LV}, where $C_2(\alpha_s)$ was calculated
up to three loops,
\begin{equation}
     C_2(\alpha_s) = 1 - {\alpha_s(Q^2) \over \pi}
          -3.58 \left({\alpha_s(Q^2) \over \pi}\right)^2
          -20.21\left({\alpha_s(Q^2) \over \pi}\right)^3
          + ... ,
\end{equation}
for three flavors. Here the renormalization scale
has been chosen to be $\mu^2=Q^2$, and so there are no
logarithms of type $\ln^k Q^2/\mu^2$ in $c_n$. Notice that
the coefficient $c_n$ is growing rapidly
with $n$ . An estimate for the fourth-order
term gives $-130(\alpha_s(Q^2)/\pi)^4$ \cite{KS}. As a matter of
fact, there are indications that $c_n$
may grow like $n!$ for large $n$ with a fixed sign!

To see the rapid growth of $c_n$, consider the contribution
from a $n$-loop bubble-chain diagram shown in Fig. 1a.
Using dimensional regularization and modified minimal
subtraction ($\overline{\rm MS}$), I find,
\begin{eqnarray}
      c_{n+1}({\rm fig.1a}) = && -{C_F \over 4\pi}({\beta_0'\over 4\pi})^{n}
             {\rm Fn}\Big[2(1-\epsilon)
             \Gamma(2-\epsilon/2) \sum^n_{k=0}
              C_n^k(-{2\over \epsilon} + \gamma)^k \nonumber \\ &&
             \times\left({6\Gamma^2(2-\epsilon/2)\Gamma(\epsilon/2)
           \over \Gamma(4-\epsilon)}\right)^{n-k}
          {\Gamma((n+1-k)\epsilon/2)\Gamma(1-(n+1-k)\epsilon/2)
              \over \Gamma(1+(n-k)\epsilon/2)
                    \Gamma(3-(n+2-k)\epsilon/2)} \Big],
\end{eqnarray}
where $C_F=4/3$ and $\beta_0' = 5-2n_f/3$ represents the
contribution to the leading-order $\beta$-function
from the one-loop gluon self-energy ($n_f$ is the number
of flavors). The symbol Fn refers to taking
finite part of the expression, which turns out to be
non-trivial. With the help of the symbolic program
MAPLE, I get the following result for the rational part,
\begin{eqnarray}
       &n \ \ \ \ \ \  & {\rm \ \ \ \ Fn[...]}    \nonumber \\
       &20 \ \ \ \ \ \  & 2.5761918 \times 10^{19}  \nonumber \\
       &21 \ \ \ \ \ \   & 5.4100006 \times 10^{20}  \nonumber  \\
       &22 \ \ \ \ \ \  & 1.1902018 \times 10^{22} .
\label{fac}
\end{eqnarray}
The irrational part that involves Reiman $\zeta$-functions
and $\pi$ is small compared with the rational part,
at least at small orders. Eq. (\ref{fac}) indicates that
the contribution from the bubble-chain diagram grows
like $\sim -5\times n!$ at large orders.

One might argue that the large coefficients from this particular
diagram might cancel with those from other
diagrams at the same order. However, indications are that
the large $n$ behavior of $c_n$ for the sum of all $n$-loop
diagrams is qualitatively similar to the bubble-chain diagram.
To see that, consider the Borel-transformed series for $C(\alpha_s)$,
\begin{equation}
     C(b) = \sum_{n=0}^{\infty} {c_{n+1}\over n!} b^n .
\label{bor1}
\end{equation}
Then $C(\alpha_s)$ can be constructed from $C(b)$ through
\begin{equation}
     C(\alpha_s) = c_0+\int^{\infty}_0 C(b) e^{-b/\alpha_s}db .
\label{bor2}
\end{equation}
Then a fixed-sign $n!$ increase is reflected by
the (renormalon) poles of $C(b)$ on the positive real axis
on the $b$ plane. From an analysis of the
infrared physics in QCD, 't Hooft argued
the existence of these poles\cite{HOO}. As I shall
explain later in more detail, it is the infrared
behavior of QCD that leads to the appearance
of the renormalon poles, and hence the
divergence of the perturbation series.
The locations of the poles at $b_n = {4\pi n /\beta_0}
\ (n = 1, 2,...)$ were identified by Parisi through a
study of the IR divergence of the corresponding
super-renormalizable
theory at $4-\epsilon$ dimension \cite{PAR}. A
more accurate description of the analytical structure
of the IR renormalons was obtained by Mueller, who showed
that the QCD radiative effects change the poles into
branch points in the complex $b$ plane\cite{MUL2}.
The leading branch point at $b_1=4\pi/\beta_0$
gives rise to a large $n$ behavior of $c_n$,
\begin{equation}
        c_{n+1} \sim n!n^\gamma \left({\beta_0\over 4\pi}\right)^n ,
\end{equation}
where $\gamma$ is related to $\beta_2$, the two-loop beta
function, and $\gamma_1$, the one-loop
anomalous dimension of the twist-four operator \cite{MUL1}.

The existence of the IR renormalons destroys the Borel summability
of the perturbation series since the integral in Eq.(\ref{bor2})
is now ill-defined. Unlike Borel-summable series, the
IR-renormalon-plagued series reflects the truly divergent
nature of the expanding quantity in perturbation theory;
the divergence has nothing to do with the choice of the expansion
parameter. Since at large $Q^2$ higher-twist terms are exponentially
small compared with the terms in the leading-twist,
and since different ways of regularizing
the Borel integral in Eq. (\ref{bor2}) produce an uncertainty
of order of $1/Q^2$, the the nature of the OPE clearly
deserves a more careful examination \cite{DAV}.

\section{Composite Operators with power divergence}

Composite operators in QCD have a special importance:
The chiral symmetry breaking is characterized
by a non-vanishing expectation value of $\bar \psi\psi$
in the vacuum. The electro-weak couplings of quarks
are through vector and axial-vector currents $\bar
\psi \gamma^\mu \psi$ and $\bar \psi \gamma^\mu \gamma_5 \psi$.
The trace anomaly of the energy-momentum tensor are proportional
to $F^2=F^{\alpha\beta}F_{\alpha\beta}$, and
the axial anomaly is proportional to $F^{\alpha\beta}\tilde
F_{\alpha\beta}$. Finally, Wilson's operator product expansion
employs myriad composite operators.

Composite operators are badly divergent in field
theory, and are not made finite through the standard
renormalization of Green's functions \cite{MUTA}.
Rather, they are renormalized individually and
have their own renormalization scale dependence.
For these reasons, composite operators are normally not
physical observables. When they do appear as a part of
a physical quantity, the renormalization scale is determined
either by physics consideration or cancelled by
the scale-dependence of other quantities.

According to the divergence behaviors of composite
operators in QCD, they can be classified into three
categories (I disregard operators with vanishing physical
matrix elements, which include BRST-exact operators and
the equations of motion operators \cite{COL}):

\begin{itemize}
\item{Finite operators: These include vector and non-singlet
axial-vector currents of quarks and their divergence like
$\bar \psi m\psi$, the energy-momentum tensor, the angular
momentum density, the topological charge density $F\tilde F$, etc.
These
operators are associated with the symmetries of the QCD lagrangian,
and are physical observables of the system.}
\item{Operators with logarithmic divergence: All twist-two
and twist-three operators have logarithmic, and only logarithmic,
divergence. Here I restrict
myself to just local operators. Let me remind the reader
that the twist of an operator is defined by the difference of the
dimension of the operator and its rank of spin in
representations of the Lorentz group. Although the notion of twist arises
from study of deep-inelastic scattering, no reference here
is needed to this application. [All operators
can be classified in terms of twist because of the Lorentz symmetry.]
The twist-two and twist-three operators have just logarithmic
divergence because their matrix elements in physical
states are dimensionless quantities. They cannot mix
under renormalization with any operators of lower dimensions.
However, they can mix among themselves, yielding a dimensionless
mixing matrix. The renormalized operators
depend on the genuine renormalization
scale $\mu^2$.}
\item{Operators with power divergence: Operators
of twist-four and higher normally have power divergence.
Exceptions include finite operators discussed
above and operators with special
symmetry properties. The simplest example of operators with
power divergence is $F^2$. One might argue that
since the energy-momentum tensor
is a finite operator and $F^2$ appears in its
trace (trace anomaly),
so $(\beta(g)/2g)F^2$ must be a finite operator. This is true
as long as one is talking about the difference of the matrix
elements in the excited states of QCD and the vacuum. Since
the vacuum energy-momentum density is not a physical
observable, $(\beta(g)/2g)F^2$ in the vacuum needs
not to be finite (the
proper normal ordering of the operator is always implied here.)}
\end{itemize}

Once an operator has quadratic or quartic or higher-order
divergence, they can mix under renormalization with lower-twist
operators. For instance, $F^2$ can mix with the trivial
operator 1, and $\bar \psi\tilde F^{\mu\nu}\gamma_\nu \gamma_5\psi$ can
mix with $\bar \psi \gamma^\mu \psi$, etc. As shall
become clear soon, these mixings have
important implications about the twist
separation in the OPE.

Let me consider the renormalization of power-divergent
operators in dimensional regularization.
Although one cannot do non-perturbative calculations
in this regularization, the coefficient functions in the
leading-twist are normally calculated in the scheme \cite{BAR}.
Since perturbative QCD does not have any mass
scale, the mixing of higher-twist operators
with lower-twist ones vanishes identically.
[Integrals of type $\int d^dk/k^m$ are taken to be zero.]
Only logarithmic divergence appears in
the matrix elements of the operators, which can be
renormalized in a standard way. Thus
it seems that composite operators can be defined up
to their logarithmic divergence and are devoid
of any power-dependent perturbative contributions.

This standard treatment of composite operators in
dimensional regularization is in fact deceptive.
In principle, infrared and ultraviolet physics are
entirely different and shall not be mixed. However,
the statement $\int d^dk/k^2=0$, for instance, has
a ridiculous implication that the ultraviolet
divergence is cancelled by the integration
in the infrared region! Not only the integration
is not reliable in the low $k^2$ region due to
non-perturbative physics, but also is finite.
Thus the implied cancellation can never
occur in reality. Actually, the
ultraviolet power divergence should not be thrown
away like this because they dictate the
correct operator mixing. Furthermore, a perturbative
calculation in the low $k^2$ region generates
IR renormalons just like the coefficient functions
discussed in the last section. Thus $\int d^dk/k^2=0$ is
mathematically convenient, but physically unwanted.

Indeed, in a study of the two-dimensional O(N) non-linear
sigma model, David found that the composite
operators with power divergence are not well-defined in
dimensional regularization \cite{DAV}.
He calculated the spin-wave condensate in
the physical vacuum using the regularization (which is possible
in $1/N$ expansion) and found that the result is not
unique. His argument for the presence of the ambiguity
in the composite operator is subtle and roughly goes like this:
When $\epsilon=4-d$
is kept finit in dimensional regularization, the higher-twist
matrix element has a series of poles on the positive real-axis
of the $\epsilon$ plane starting from $\epsilon$ =
the degree of divergence of the operator.
The poles are present because at these $\epsilon$,
one cannot distinguish a higher-twist contribution from
a lower-twist one. These poles make the limit of
$\epsilon \to 0$ ambiguous, depending on whether one
takes Im$\epsilon>$ or $<$ 0.

The interpretation of David's ambiguity is
that composite operators in dimensional
regularization have IR renormalons.
They were not there originally, but
appear when the scheme insists to treat the contribution
in the soft momentum region in perturbation theory.
Those perturbative soft
contributions are needed to cancel the ultraviolet power
divergence because the relations like $\int d^dk/k^m=0$
are used. Consequently the power
divergence is killed with
a price: emergence of IR renormalon singularities.
It is these singularities that cause the ambiguity
in David's calculation. [An explicit example of the
renormalons in matrix elements is
provided in the next section in Eq. (20)]
Thus in dimensional regularization,
in addition to subtract the logarithmic
divergence of an operator,
one needs to specify how the IR renormalons are regulated.

What if one uses other regularization
schemes, such as the scheme with a momentum cut-off?
In a cut-off scheme, composite operators are finite,
as long as the cut-off is kept finite. The operators are
also free of any IR renormalon singularities.
Their physical matrix elements depend on two
scales: the non-perturbative scale generated
from the breaking of the scale symmetry in the
non-perturbative
QCD \cite{CRE} and the perturbative scale that is
served as a cut-off for both the log and power divergences.
If the degree of divergence of an operator is $d_O$,
its matrix element goes like the $d_O$ power of
the cut-off when the cut-off approaches infinity.
In light of dimensional regularization, it is
tempting to define a {\it pure} non-perturbative matrix
element by subtracting off the corresponding perturbative
matrix element. Although the cut-off dependence now disappears,
the IR renormalon singularities are introduced through
perturbative matrix elements. Thus a pure non-perturbative
matrix element is again not well-defined.

To summarize the above discussion,
operators with power divergence
cannot be specified uniquely after subtracting
their logarithmic divergence in dimensional regularization.
One must specify in addition how the IR renormalons
are regularized, or, equivalently, how the limit
$\epsilon \to 0$ is taken. In a cut-off regularization,
the operators are cut-off dependent, independent
of how the logarithmic divergence is regulated. For example,
$\beta(g)/2gF^2$ has no logarithmic divergence at all
but still is cut-off dependent.
One important consequence of this is that
there are no physically
well-defined higher-twist contributions in the OPE.

Since the physical quantity that is
expanded in the OPE is well-defined, the ambiguity in
defining higher-twist operators
must be correlated with
the infrared renormalon
singularities in the coefficient functions.
In the 2D non-linear sigma
model, David showed that the different
choices for regularizing the Borel integral
and defining the composite operators
give the same final result. Thus despite the existence
of arbitrariness in defining the contributions
of different twists, the spirit of OPE
is unspoiled.

In fact, the situation with the twist separation
is analogous to the factorization of
collinear singularities in the OPE \cite{MUTA}.
There the leading-twist
contribution in the OPE can be written as a product of the
coefficient functions and the hadron matrix elements.
The coefficient functions depend on the factorization
scale because they are obtained by subtracting off
collinear divergence in a single
parton scattering. The matrix elements are scale-dependent
because the local operators are
renormalized operators disposed of ultraviolet divergence.
The factorization scale controls how
much physics is considered to be perturbative and how much is
non-perturbative. Their product is
scale-independent.

\section{operator product expansion in a cut-off scheme}

According to the previous two sections,
the traditional recipe for the OPE is at best formal:
The coefficient functions are intrinsically
divergent and higher-twist operators are ill-defined
after removing the logarithmic divergence.
However, both problems
have a common origin---the IR renormalons. The existence of
the renormalons makes the notion of twist separation
procedure-dependent, however, specifics
cancel in the physical sum. Specifically,
one needs to define the limit $\epsilon \to 0$
in higher-twist operators consistent with handling
of the Borel integral for the coefficient functions \cite{DAV}.
However, the approach is hardly useful in practical
applications. First, no one knows how to calculate
the non-perturbative matrix elements of higher-twist
operators in dimensional regularization in QCD.
Second, despite tremendous progress made recently in calculating
the QCD perturbative series to higher orders \cite{KS}, it remains
a formidable task to calculate the exact behavior of a
infrared renormalon series. Lastly, even all the above
are possible, the regularization does not seem
to have clear physical motivation.
[The author noticed a number of
other proposals in the literature for
regularization the infrared renormalons \cite{MUL1,GRU}.]

To my opinion, a better approach to the OPE, which was
advocated by the former ITEP group \cite{SVZ,NOV},
and was first taken seriously
by Mueller \cite{MUL2}, is to consider the
twist expansion entirely from the point
of view of Feynman diagrams.
Imagine a set of Feynman diagrams completely define
a physical cross section or a correlation function in QCD.
Often in accompany with a high-energy subdiagram, there are
soft subprocesses mediated by low-energy quarks and gluons,
which result in soft (infrared) and collinear (mass)
divergence in perturbative calculations \cite{MUTA}.
In many physical observables, the soft divergence
cancel according to equivalent of the Kinoshita-Lee-Nauenburg
theorem \cite{KLN}, and the collinear divergence
can be factorized into hadron matrix elements,
which in the end are replaced with
proper non-perturbative ones \cite{CSS}. The remaining part
is calculable as a perturbation series in the strong
coupling constant $\alpha_s$ and is called infrared safe.
The classical examples of the
perturbative-calculable quantities include
the total rate for $e^+e^-\to$ hadron, the
coefficients in Wilson's operator product expansion, and
radiative corrections to jet production, Drell-Yan
cross section, etc.

However, it was pointed out by Shifman, Vainshtein,
and Zakharov (SVZ) that using ``infrared-safe"
as a criteria to decide
a perturbative calculation is inadequate \cite{SVZ}.
They argued, in the case of current correlation
functions in the vacuum, that the part
of Feynman diagrams involving low-virtuality
quarks and gluons cannot be calculated perturbatively despite
such a calculation contains no infrared divergence and the contribution
might be small. A legitimate calculation for the part must be done
non-perturbatively, by using, in their case, various
vacuum condensates. These vacuum condensates correspond
to the higher-twist terms in Wilson's OPE.
Thus, the role of higher-twist contributions is
to take into account properly the residual
soft contributions in Feynman diagrams after the
KLN type of cancellation.

Thus to devise a twist expansion,
one can start by separating the hard and soft
contributions in Feynman diagrams. This is precisely
what Mueller did for vacuum correlation
functions \cite{MUL2}. In this way, he obtained
coefficient functions that are
entirely perturbative and free of IR renormalons,
and the higher-twist operators that are free
of ultraviolet divergence due to an explicit cut-off.
Both contributions now
depends on the twist-separation scale
but the sum does not. Notice the separation
scale here plays the same role as the prescriptions of
regularizing the higher-twist operators in
David's paper. In this section, I follow Mueller's
approach in Ref. \cite{MUL2} and attempt to
construct a twist expansion for the Bjorken sum rule.
To motivate the approach, let me come back to
the physical origin of the renormalon singularities in the
perturbation series.

The appearance of $n!$ in the
coefficient functions is due to loop integrations in small
momentum regions in Feynman diagrams \cite{MUL2,ZAK}. In fact,
consider a one-loop Feynman diagram
with the gluon momentum $k$ (chosen to be the loop momentum).
Replace the gluon coupling $\alpha_s(Q^2)$ (again, $\mu^2=Q^2$)
with a running coupling $\alpha_s(k^2)$,
\begin{equation}
      \alpha_s(k^2) = {\alpha_s(Q^2)\over 1
  + \alpha_s(Q^2){\beta_0 \over 4\pi}.
          \ln{k^2\over Q^2}}
\end{equation}
Then the small $k^2$ integration is questionable because where
the perturbative expression for $\alpha_s(k^2)$ breaks down.
[Notice the existence of the familar
Landau ghost pole at $k^2=Q^2\exp(-\alpha_s(Q^2)\beta_0/(4\pi))$.]
On the other hand, when expanding $\alpha_s(k^2)$ in terms
of $\alpha_s(Q^2)$,
\begin{equation}
     \alpha_s(k^2) = \alpha_s(Q^2)\sum_n (-1)^n
       ({\alpha_s(Q^2)\beta_0 \over 4\pi})^n \ln^n({Q^2\over k^2}),
\end{equation}
the $n$-th term represents a legitimate partial contribution
from the $n$-th order Feynman diagrams.
When it is inserted in the loop integral, the repeated
partial integrations give $n!$. A simple
examination reveals that most of the contribution
comes from the region of the integration
$k^2 \sim Q^2 e^{-n}$, where the logarithms become
large.
Therefore the divergence of the perturbation theory is caused
by illicit perturbative calculations in the soft
regions where perturbative quarks and gluons are ill-defined. Although
order by order, perturbative contributions
from these regions are finite, they diverge as $n\to \infty$
irrespetive the value of $Q^2$.
Thus, a physically motivated OPE shall incorporate
the perturbative soft (infrared) contributions
in the higher-twist terms.

To identify such contributions in the Bjorken sum rule,
consider the one-loop contribution
to the coefficient function shown
in Fig. 2. The one-loop integrations are free of
ultraviolet divergence after the
standard QCD renormalization. The external quark momentum
is set to zero after the collinear singularity
in the box diagram (Fig. 2d) is subtracted by
the one-loop matrix element of the axial current \cite{BAR}.
Now focus on contributions
from the region where the gluon momentum is soft
($k^2$ is small). In this region, one can expand the
integrand in $k^2/ Q^2$ and keep the
leading-order contribution. Applying an upper cut-off
$\Lambda^2$ on the gluon virtuality, I find the perturbative
part of $c_1$ from the soft region,
\begin{eqnarray}
        \Delta c_1^{\rm soft \ pert.} &=& {8g^2\over 3Q^2}C_F
        \int^{\Lambda^2}_0
                {d^4k \over (2\pi)^4}{-i\over k^2}  \cr
                 &=& -{8\over 9}{\alpha_s(Q^2) \over \pi}
        {\Lambda^2\over Q^2},
\label{soft1}
\end{eqnarray}
where in the second line
have rotated the integration to the Euclidean
space. The factor 8/3 in the first line is the sum of
2, 1, and $-$1/3, coming from  diagrams a), b), and c) in Fig. 2,
respectively.
The cut-off $\Lambda^2$ should be on the order of
0.5 to 1 GeV$^2$, above which perturbative calculation
is justifiable. Clearly,
the contribution in Eq. (\ref{soft1}) must subtracted from
$c_1$ and the corresponding non-perturbative
contribution should be added as a higher-twist
contribution.

To find the correct non-perturbative contribution,
I re-examine the one loop diagrams in Fig. 2.
When the gluons are soft, we cannot contract
the field operators in the perturbative vacuum,
instead we must keep them and evaluate the
matrix elements in the non-perturbative zero-momentum quark state.
[The same thing
can be said of any quark lines that carry
soft momenta.] For example, the diagrams (b) and (c) in Fig. 2
contribute to the quark Compton amplitude,
\begin{equation}
      T^{\mu\nu}_q = \int {d^4k\over (2\pi)^4}
              {d^4k_1\over (2\pi)^4} H^{\mu\nu\alpha}(k,k_1)
             S_\alpha(k,k_1) ,
\end{equation}
where $k$ and $k_1$ are quark and gluon momenta, respectively, with
Euclidean $k_1^2$ restricted to $\le \Lambda^2$.
$H^{\mu\nu\alpha}$ represents part of the diagrams above the dashed
lines and is perturbative. $S_\alpha$ represents that below
and can be obtained from the first-order calculation of
\begin{equation}
        S_\alpha(k, k_1)
      = \int d^4\xi d^4\xi_1 e^{ik\cdot \xi}e^{ik_1\cdot \xi_1}
        \langle q(0)|{\rm T}\bar \psi(\xi)A_\alpha(\xi_1)\psi(0)|q(0)\rangle,
\end{equation}
where $\psi$ and $A^{\mu}$ are quark and gluon field operators and
$|q(0)\rangle$ is a zero-momentum quark state.
Thus the one-loop diagrams arise from a
perturbative expansion of the zero-momentum-quark
wave function, which of course is incorrect
as the gluon momentum becomes soft. However, so long as
one refrains from such a perturbative expansion, one gets the correct
contribution from the soft-gluon region.

To proceed, I use the technique of the
collinear expansion as explained in full details
in Ref. \cite{EFP}. The result is,
\begin{equation}
 T^{\mu\nu}_q = 2i\epsilon^{\mu\nu\alpha\beta}
            {8q_\alpha \over 9Q^4} \langle q(0)|O_{4\beta}^{\rm cut}
             |q(0) \rangle
\end{equation}
where,
\begin{eqnarray}
       O_{4\beta}^{\rm cut}
        &=& \int_{k_1^2,k_2^2<\Lambda^2}
        {d^4k_1\over (2\pi)^4}{d^4k_2 \over (2\pi)^4}
       d^4\xi_1 d^4\xi_2 e^{ik_1\cdot \xi_1}e^{ik_2\cdot \xi_2}
    \nonumber \\
       && \times\, \bar \psi(0)\left[\gamma_\beta\gamma_5
       i{ D}_\perp(\xi_1) \cdot
      i{D}_\perp(\xi_2) + i\epsilon^{\alpha\eta\gamma\delta}
       p_\gamma n_\delta \gamma_\beta iD_\alpha(\xi_1)
       iD_\eta(\xi_2)\right] \psi(0)
\end{eqnarray}
where $p$ and $n$ are light-like vectors that are introduced
as a part of the coordinate basis.  Clearly, as $\Lambda^2 \to \infty$,
$O_{4\beta}^{\rm cut} \to O_{4\beta}$, the local twist-four
operator appearing in Eq. (\ref{ope}). For
a fixed $\Lambda^2$, $O_{4\beta}^{\rm cut}$ is a non-local operator.
It is simple to show that at the one-loop order the
perturbative calculation of the twist-four matrix
element reproduces the soft perturbative
contribution to the coefficient function in
Eq. (\ref{soft1}).

The above result tells us several things about the operator
product expansion in Eq. (5). First, the higher-twist
operator has an upper cut-off in the Fourier
components for the gluon fields, and thus is finite.
To fully include the non-perturbative effect,
the cut-off $\Lambda^2$
must be larger than any non-perturbative
scale, such as $\Lambda_{\rm QCD}$. Second, when quarks
and gluons have virtuality greater than
$\Lambda^2$, their effects are
perturbative and are included in the
coefficient functions. The separation of the two effects
are scale-dependent, but the sum is not.
Lastly, although I discussed only one-loop
diagrams with one soft-gluon, the non-perturbative
twist-four contribution takes care of all Feynman
diagrams with one soft-gluon. Diagrams with one soft-quark
line do not contribute at the twist-four level \cite{JI2}.
Diagrams with two soft-gluon or quark lines are
related to the twist-six terms, which are beyond the
scope of this paper.

Clearly, when taking $O_{4\beta}^{\rm cut}$ as
a definition of higher-twist operator,
the corresponding coefficient function
in Eq. (\ref{ope}) shall be modified,
\begin{equation}
      C_2(\alpha_s) = C_{2\rm pert}(\alpha_s)
             + {8\over 9}C_4(\alpha_s)
        {\langle O_{4\beta}^{\rm cut}\rangle_{\rm pert}
\over Q^2}
\label{result}
\end{equation}
where $C_{2\rm pert}(\alpha_s)$ the usual perturbation series
in Eq. (\ref{pert}), $C_4(\alpha_s)$ is the coefficient function
for the twist-four operator.
The perturbative matrix element of $O_{4\mu}^{\rm cut}$
is to be evaluated in a zero-momentum quark state
and is expected to have IR renormalons.
Since soft contributions are subtracted in Eq. (\ref{result}),
we expect that $C_2(\alpha_s)$ is free of the
leading IR renormalon singularity. However, the
renormalon poles at $4\pi n/\beta_0$ with $n\ge 2$
are still present.
To subtract them, one must consider
the two-loop contributions to $C_{2\rm pert}(\alpha_s)$
and the twist-six operators.

To see that the perturbative matrix element of
$O^{\rm cut}_{4\beta}$ has IR renormalons which cancels
the leading IR renormalon in $C_{2\rm pert}(\alpha_s)$,
I consider again the bubble-chain
diagram shown in Fig. 1a. The corresponding
contribution to the matrix element is shown in Fig. 1b.
A simple calculation shows,
\begin{eqnarray}
      \langle O_{4\beta}^{\rm cut} \rangle_{\rm pert}({\rm Fig.1b})
     & = & {\alpha_s(Q^2)\over 4\pi}C_F
     \left({\beta_0'\alpha_s(Q^2)\over 4\pi}\right)^n
     {\rm Fn}\Big[3\Lambda^2 \sum_{k=0}^nC_n^k {1\over 1-\epsilon(n-k)/2}
      \nonumber \\ &&
        \left[{6\Gamma(2-\epsilon/2)^2\Gamma(\epsilon/2)
           \over \Gamma(4-\epsilon)}\right]^{n-k}
           \left(-{2\over \epsilon} + \gamma\right)^k
         \left({Q^2\over \Lambda^2}\right)^{\epsilon(n-k)/2} \Big]
\end{eqnarray}
Thus the subtraction is both $\Lambda^2$ and $Q^2$ dependent. If
I fix $\Lambda^2$ at 1 GeV$^2$, the subtraction is maximum
at $Q^2=1$ GeV$^2$. As $Q^2$ increase, the subtraction
becomes smaller, indicating less contributions in $c_n$
from the non-perturbative region. For large $n$, the
subtraction is a exceedingly slow-varying function of $Q^2$. In fact,
defining $y=\Lambda^2/Q^2$, I find at $n=20$,
\begin{eqnarray}
   &y     \ \ \     &{\ \ \ 2{\rm Fn[...]}/ (3Q^2)}   \nonumber \\
  &0.1   \  \ \     &2.5761951\times 10^{19} \nonumber \\
  &0.01  \  \ \     &2.5761877\times 10^{19} \nonumber \\
  &0.001  \  \ \    &2.5755872\times 10^{19}
\end{eqnarray}
Comparing with the result from Eq. (8),
the subtraction is accurate up to one part per million for $y=0.1$.
As $n\to \infty$,
the large perturbative contributions to $C_{2\rm pert}$
are entirely subtracted, independent of $Q^2$ and $\Lambda^2$.

\section{Twist-four contribution in point-splitting regularization}

The cut-off scheme discussed in the previous
section best illustrates the goal of the OPE:
separating the perturbative and
soft (infrared) contributions in a process. In
analogy with the factorization
of the collinear singularities at each twist,
the twist separation is not unique.
In the cut-off scheme, this is shown by
dependence of higher-twist
operators on the cut-off scale $\Lambda$.
The size of $\Lambda$ must be large enough so that
all non-perturbative physics is covered by
higher-twist matrix elements, and at the same time
shall be small enough that the pure perturbative
physics is included in the coefficient functions of
the leading-twist terms. In QCD, one believes
that such scale exists around 1 GeV$^2$.

In view of what twist-expansion accomplishes,
I argue that there exists a large
degree of arbitrariness in choosing a
cut-off version of the twist-four
operator $O_{4\mu}$: The specific
cut-off scheme that I discussed before is only one example.
As long as it fully incorporates the non-perturbative
physics, as long as it approaches $O_{4\mu}$
when cut-off is let go to infinity, any
finite, non-local operator with right dimension and
quantum numbers is a good twist-four operator.
Once a choice is made, the coefficient function of
the twist-two term can be calculated through
applying the OPE to, say, a perturbative
zero-momentum quark state. Thus the computation of the
coefficient functions goes hand in hand with the
choices of the higher-twist operators.

The freedom of twist-separation is very useful
in practical applications. The twist-four operator
in the cut-off scheme is quite complicated,
not suitable for non-perturbative methods.
In what follows, I seek a definition of
the twist-four operator convenient for lattice
calculations, because the lattice is the only
practical means at present for a non-perturbative
calculation. Following the definition, I
calculate the corresponding twist-two
coefficient function. The approach here can be
generalized straightforwardly to the application
of the OPE to the QCD sum rules, which will be
discussed in a separation publication \cite{JI}.

I choose to define composite operators in Euclidean
space, in which the lattice realization is usually made.
Imagine the space-time is discretized into a
lattice with hypercubes; the lattice spacing is
$\pi /\Lambda$. The quark fields
are assigned at every lattice point and
gluon fields are represented by gauge links with
\begin{equation}
        U_\mu(n) = {\cal P}e^{-ig(\pi/\Lambda)\int^1_0
                  A_\mu(n+x\hat \mu)dx}
\end{equation}
representing a link, starting at the discrete point $n$
and ending at a neighbouring lattice point in the
direction $\hat \mu$. ${\cal P}$ is the path-ordering
operator. I define an operator on lattice,
\begin{equation}
O_{4\alpha}(\Lambda^2, n) = i{\Lambda^2 \over 4\pi^2}
              \epsilon_{\alpha\beta\mu\nu}
            \bar \psi(n)[U_\mu(n)U_\nu(n+\hat \mu)
             - U_\nu(n)U_\mu(n+\hat \nu)]\gamma_\beta
           \psi(n+\hat \mu+\hat \nu) + {\rm h.c.}
\label{o4l}
\end{equation}
where h.c. represents the hermitian conjugation term. The
definitions of various Euclidean variables follow
those of Rothe \cite{ROT}.
It is simple to show that in the limit of $\Lambda^2 \to
\infty$, the operator goes to $O_{4\alpha}$.
Under the standard lattice gauge transformation,
\begin{eqnarray}
       \psi(n) &&\rightarrow u(n)\psi(n) \nonumber \\
        U_\mu(n)&& \rightarrow u(n) U_\mu(n) u^\dagger(n+\hat\mu)
\end{eqnarray}
where $u(n)$ is an element of the color SU(3) group,
the operator is gauge invariant. Since the gluon fields are
integrated along a straight path in a distance of
$\pi/\Lambda$, the large momentum components of the fields
have effectively been cut off---the operator is finite.

A few clarification is in order. First of all,
I have said nothing about lattice QCD so far. I
use the lattice as a tool to motivate a non-local,
gauge invariant twist-four operator. Thus
the actual quark and gluon fields and the location $n^\mu$ of
the plaquets can still be regarded as continuous.
[The method of defining finite operators
is very similar to the well-known Schwinger's method
of splitting the fermion fields in a current.]
Second, the operator is non-local in Euclidean time,
thus one has to rotate all loop integrals to
Euclidean space in a perturbative calculation.
Lastly, the operator has no complete
Lorentz symmetry. However, the violation occurs at the scale
of $\Lambda$ around which physics is perturbative.
When its matrix elements are used in conjunction
with the coefficient function defined below, the
symmetry breaking effects largely cancel.

With the above definition of the twist-four operator,
one can calculate the non-perturbative matrix
element in a polarized nucleon state in lattice QCD.
For instance, by choosing a series of lattice
spacings $a_n = \pi/(n\Lambda)$ ($n=1,2,...$),
one can construct the operator from lattice
variables almost trivially.
By comparing a series of calculations
with different $n$, one can extract the matrix element
in the continuum limit. [Note that the scale $\Lambda$ in
the operator shall not be changed as the lattice spacing
$a$ approaches zero. The presence of this scale
ensures the existence of the finite matrix element
in the limit.] The lattice calculation
is currently been persued by the MIT lattice group \cite{MIT}.

What remains is to calculate the corresponding
coefficient function of the twist-two operator.
As usual, I sandwich the OPE in a zero-momentum
quark state and calculate everything using dimensional
regularization, including the perturbative matrix element
of the twist-four operator. At one-loop level, the
matrix element is,
\begin{eqnarray}
      \langle q(0)|O_4^\mu(\Lambda^2)|q(0) \rangle
          &=& g^2C_F{\Lambda^2\over \pi^2} \sum_{\mu\nu}\int
           \sin \left({k_\mu\pi\over \Lambda}\right)
           \sin \left({k_\nu\pi\over \Lambda}\right)
              {k_\mu\over k_\nu} {1\over k^4}
             {d^4k\over (2\pi)^4} \,\bar u\gamma^\mu\gamma^5 u  \nonumber \\
          & = & {2\alpha_s\Lambda^2 \over \pi^2}\bar u\gamma^\mu\gamma^5 u .
\end{eqnarray}
The calculation is done with Feynman rules in Euclidean space.
 From this we get the one-loop contribution
to the coefficient function,
\begin{equation}
     c_1 = -{1\over \pi}(1-{16\Lambda^2\over 9\pi Q^2})
\end{equation}
The power term here is quite close to the power term in the
cut-off scheme in the previous section.
I emphasize again that this coefficient function
must be used together with the specific
twist-four operator in Eq. (\ref{o4l}).

Another way to define the twist-four operator
is to let the non-locality of the operator in Eq.(\ref{o4l})
$\Lambda$ be the same as the lattice spacing $a$,
approaching zero in the continuum limit.
Then the lattice matrix element
diverges like $1/a^2$. A finite twist-four
operator can be defined by subtracting from the
original operator its perturbative matrix
element on the lattice, with
a low momentum cut-off in loop integrals.
Thus the twist-four contribution is again
free of the IR renormalons but still depends on the
infrared cut-off. Unfortunately, it is difficult to
do perturbative calculations with a
nucleon state. However, the idea is clearly
applicable to vacuum condensates in the QCD
sum rules\cite{LAT,JI}.

\section{comments and conclusions}

According to the discussions in the previous
sections, the leading-twist contribution to
the Bjorken sum rule can still be regarded as
a power series in $\alpha_s(Q^2)$. However, the
coefficient in each order is now not just
a numerical number; it contains the subtraction of
the soft contributions that are represented by a power
series in $\Lambda^2/Q^2$. The most important
soft-subtractions are those nominally suppressed
by one power of $1/Q^2$. As the order of
the perturbation $n\rightarrow \infty$,
the subtraction grows like $n!$,
independent of the value of $Q^2$ and
the subtraction scale $\Lambda^2$.
Therefore, the series in $\alpha_s$
would converge if other renormalons
singularities and sources of divergent factors
did not exit.
[It is known that ultraviolet renormalons and
the singularities in the Borel plane caused by
instanton-anti-instanton pairs still
make the series asymtotic, but I ignore them here.]

Thus to the accuracy of $1/Q^2$, one can successively
take into account the QCD
corrections in the following way.
First calculate the coefficient function $C_2(\alpha_s,
\Lambda^2/Q^2)$ in powers of $\alpha_s$. At each other,
the $1/Q^2$ subtraction must be calculated explicitly.
At lower orders, the subtraction is smaller than
the pure loop contributions. However,
the former must be kept to maintain the accuracy.
As $n$ increases, both the loop contributions and
subtractions decrease initially, and their
difference also decreases. If the difference
becomes the size of the non-perturbative
$1/Q^2$ terms, the latter must be included.
With the further increase of $n$, both the perturbative
contributions and their subtractions
start to grow. However, the cancellation
now becomes more complete and
the residual remains small. The first few terms
of the Bjorken sum rule is,
\begin{eqnarray}
  \int^1_0(g_1^p(x, Q^2)-g_1^n(x, Q^2))dx
& = & {g_A\over 6}
\left(1-{\alpha_s(Q^2)\over \pi}(1-{16\Lambda^2\over 9\pi Q^2})
      - 3.58\left({\alpha_s\over\pi}\right)^2(1-\eta{\Lambda^2\over Q^2})-...
     \right) \nonumber \\  && -
    {4\over 27}{\langle O^\mu_4(\Lambda^2) \rangle \over Q^2}+ ...
\end{eqnarray}
The soft contribution in the two-loop diagrams, denoted by $\eta$, needs
to be calculated.

The non-perturbative twist-four contributions
can be calculated on lattice, as explained in the
last section. In the previous studies, the
twist-four matrix elements have been evaluated
in the MIT bag model and the QCD sum rule method.
The bag model calculation
has no explicit cut-off dependence \cite{JU},
which is a common problem for model estimates.
Empirically, one may regard the cut-off being
somewhere from 0.5 to 0.8 GeV$^2$ \cite{JR}.
The QCD sum rule calculations for the higher-twist
matrix elements are clearly questionable
\cite{BAL}. As I explained in sec. III,
the operators used in these calculations
are not well-defined.

The IR renormalons exist in all perturbation
series in QCD. The issue is closed related to the
non-perturbative power corrections.
Unlike the case of QED, the power corrections are
an important subject in perturbative QCD. For
instance, a precision determination of the
strong coupling constant clearly needs
a correct understanding of the higher-twist effects.
In fact, already at present,
the effects become the dominating theoretical
uncertainty in some of the data analysis \cite{HIN}. The
framework described in this paper provides one way
get the problem under control.

To summarize, we have studied in this paper
the IR renormalons
and $1/Q^2$ corrections in deep-inelastic sum rules.
The result shows that the non-perturbative
$1/Q^2$ corrections depend on specific
choices of the regularized higher-twist
operators. I have proposed to define a
twist-four operator for the Bjorken sum rule,
which is suitable for lattice calculations.
The coefficient functions
for the leading-terms can only be calculated
after the higher-twist operators are defined.
Although most of the discussions in the paper
are made for the Bjorken sum rule,
all the results are clearly
applicable to all deep-inelastic sum rules.

I wish to thank A. Mueller for many useful discussions,
comments, and suggestions. Discussions with
K. Johnson, W. Lin, J. Negele, B. Schreiber, U. Wiese
are also acknowledged.

\begin{figure}
\caption{a). A bubble-chain diagram for the coefficient
function of the Bjorken sum rule. b). The corresponding
matrix element of the twist-four operator.}
\label{fig1}
\caption{The complete one-loop Feynman diagram for
the the coefficient
function of the Bjorken sum rule.}
\label{fig2}

\end{figure}
\end{document}